\documentclass[conference]{IEEEtran}
\IEEEoverridecommandlockouts

\usepackage[numbers, sort]{natbib}
\usepackage{amsmath,amssymb,amsfonts}
\usepackage{algorithmic}
\usepackage{graphicx}
\usepackage{textcomp}
\usepackage{colortbl}
\usepackage[dvipsnames]{xcolor}
\usepackage{multirow}
\usepackage{subcaption}
\usepackage{xspace}

\usepackage[hyphens]{url}
\usepackage{soul}
\NewDocumentCommand{\sotwo}{O{red}O{black}+m}
    {%
        \begingroup
        \setulcolor{#1}%
        \setul{-.5ex}{1pt}%
        \def\SOUL@uleverysyllable{%
            \rlap{%
                \color{#2}\the\SOUL@syllable
                \SOUL@setkern\SOUL@charkern}%
            \SOUL@ulunderline{%
                \phantom{\the\SOUL@syllable}}%
        }%
        \ul{#3}%
        \endgroup
    }

\definecolor{lightblue}{RGB}{207,212,220}
\definecolor{echocolor}{RGB}{214, 126,44}

\definecolor{mikecolor}{RGB}{255,140,0}

\definecolor{czcolor}{RGB}{102,204,255}

\definecolor{alcolor}{RGB}{143, 0, 255}

\usepackage[utf8]{inputenc}
\usepackage[most]{tcolorbox} 

\newcommand{\smallsection}[1]{\noindent\textbf{#1.}}

\def\BibTeX{{\rm B\kern-.05em{\sc i\kern-.025em b}\kern-.08em
    T\kern-.1667em\lower.7ex\hbox{E}\kern-.125emX}}

\tcbset{
  rqlabel/.style={
    colback=white,
    colframe=black,
    boxrule=0.7pt,
    arc=3mm,
    left=1mm,
    right=1mm,
    top=1mm,
    bottom=1mm,
    boxsep=2pt,
    fonttitle=\bfseries,
  }
}

\newtcolorbox{rqbox}[2][]{colback=red!5!white,
colframe=gray!60!black,fonttitle=\bfseries,
colbacktitle=white!70!black,enhanced,
coltitle=black,
left=4pt,right=4pt,top=4pt,bottom=2pt,before skip=5pt,after skip=5pt,
attach boxed title to top left={yshift=-3mm,yshifttext=-1mm,xshift=4mm},
title={#2},#1}

\makeatletter
\newcommand{\linebreakand}{%
  \end{@IEEEauthorhalign}
  \hfill\mbox{}\par
  \mbox{}\hfill\begin{@IEEEauthorhalign}
}
\makeatother

\newcommand{\rqonetext}{%
What tools, tasks, and usage patterns associated with \emph{GenAI} are evident in \emph{GenAI}-related issue discussions?
}

\newcommand{\rqtwotext}{%
What challenges 
do developers report in \emph{GenAI}-related issue discussions?
}

\newcommand{\rqthreetext}{%
How do the characteristics and challenges 
of \emph{GenAI}-related issues differ from those observed in \emph{TradAI} and \emph{NonAI} issue discussions?
}

\begin{document}

\title{An Empirical Study of GenAI Adoption in Open-Source Game Development: 
\\ Tools, Tasks, and Developer Challenges}

\author{%
\IEEEauthorblockN{Xiang Echo Chen}
\IEEEauthorblockA{\textit{David R. Cheriton School of Computer Science} \\
\textit{University of Waterloo}\\
Waterloo, Canada \\
x784chen@uwaterloo.ca}%
\and
\IEEEauthorblockN{Wenhan Zhu}
\IEEEauthorblockA{\textit{ACM Member}\\
Waterloo, Canada \\
wenhanzhu1@acm.org}
\and
\IEEEauthorblockN{Guoshuai Albert Shi}
\IEEEauthorblockA{\textit{David R. Cheriton School of Computer Science} \\
\textit{University of Waterloo}\\
Waterloo, Canada \\
albert.shi@uwaterloo.ca}%
\linebreakand
\IEEEauthorblockN{Michael W. Godfrey}
\IEEEauthorblockA{\textit{David R. Cheriton School of Computer Science} \\
\textit{University of Waterloo}\\
Waterloo, Canada \\
migod@uwaterloo.ca}%
}

\maketitle

\begin{abstract}
The growing capabilities of generative AI (GenAI) have begun to reshape how games are designed and developed, offering new tools for content creation, gameplay simulation, and design ideation. While prior research has explored traditional uses of AI in games, such as controlling agents or generating procedural content. There is limited empirical understanding of how GenAI is adopted by developers in real-world contexts, especially within the open-source community. This study aims to explore how GenAI technologies are discussed, adopted, and integrated into open-source game development by analyzing issue discussions on GitHub.
We investigate the tools, tasks, and challenges associated with GenAI by comparing GenAI-related issues to those involving traditional AI (TradAI) and NonAI topics. 
Our goal is to uncover how GenAI differs from other approaches in terms of usage patterns, developer concerns, and integration practices.
To address this objective, we construct a dataset of open-source game repositories that discuss AI-related topics. 
We apply open card sorting and thematic analysis to a stratified sample of GitHub issues, labelling each by type and content. 
These annotations enable comparative analysis across GenAI, TradAI, and NonAI groups, and provide insight into how GenAI is shaping the workflows and pain points of open-source game developers.
\end{abstract}


\section{Introduction}
The global game industry has become a dominant force in entertainment, generating over \$200 billion and surpassing the combined revenues of film, television, and music~\cite{GGG2024}.
While computer video games (hereafter, \textit{games}) are a thriving part of the software industry, they have been largely overlooked by software engineering researchers until recent years~\cite{pascarella2018video}. 
Games are not merely software products; they also represent creative works, cultural expressions, and, in some cases, forms of art~\cite{engstrom2018game}.
Their development requires specialized skills~\cite{tschang2007balancing} and differs significantly from other software domains~\cite{kanode2009software,murphy2014cowboys}. These interdisciplinary characteristics make game development a valuable setting to examine how emerging technologies such as GenAI are reshaping creative and technical workflows.

Games and Artificial Intelligence (AI) have shared a long and evolving history.
Traditionally, AI has been used to control game agents, support content creation, and simulate or analyze player behaviour~\cite{yannakakis2018artificial}.
Procedural Content Generation (PCG)~\cite{shaker2016procedural} plays a key role in game development by enabling the automatic creation of assets such as levels, maps, and narratives.
More recently, advances in generative AI (GenAI) have further expanded these capabilities, offering new opportunities for automatically producing a much wider range of game content and developer resources, and showing promising applications in areas such as gameplay ideation~\cite{kanervisto2025world} and generative agents~\cite{park2023generative}.

Despite this momentum, there is limited exploratory understanding of how developers use or struggle with GenAI tools in real-world game projects, especially in open-source environments where workflows and team dynamics differ from commercial development.
While prior research has examined the role of traditional AI in games and the characteristics of open-source game development~\cite{pascarella2018video,yannakakis2018artificial,scacchi2004free}, 
no study has systematically explored how GenAI is integrated into game development workflows, what tasks it supports, or how it compares to other AI techniques.

To address this gap, we propose an empirical study of issue discussions from open-source game repositories on GitHub.
We focus on issue threads because they capture real-time developer reflections on tool use, feature implementation, and integration challenges~\cite{bissyande2013got,tan2020first}. Beyond serving as task trackers, issues often include coordination, brainstorming, and technical reasoning, making them a valuable lens for examining how developers engage with GenAI tools in practice.
While our broader goal is to understand GenAI’s role in game development, open-source projects offer transparent access to real-world development activity, providing early insight into how developers coordinate, reflect, and solve problems involving GenAI tools.

We are planning to build a dataset of AI-related game repositories, apply structured labelling through open card sorting and thematic analysis, and compare \emph{GenAI}, \emph{TradAI}, and \emph{NonAI} issues across multiple dimensions. 

Our findings will serve as a starting point and offer early empirical insights into the opportunities and limitations of GenAI in game workflows, providing guidance for researchers, tool designers, and game developers seeking to better support GenAI adoption.

\section{Background and Related Work}

\subsection{Game Development}

Game development poses distinct challenges, including asset integration, frequent design iteration, and difficulty evaluating subjective qualities like engagement or fun~\cite{murphy2014cowboys, pascarella2018video}.
This section focuses on two areas of game development research: (1) AI in game development, and (2) open-source game development.

\smallsection{AI in Game Development}
Artificial intelligence has been used in games for decades. Even before AI was formally recognized as a field, pioneers such as Alan Turing explored game-playing programs to test machine intelligence, including his reinvention of the Minimax algorithm for Chess~\cite{turing1953digital}. 
Other milestones include A.~S.~Douglas’s 1952 Tic-Tac-Toe program, developed as part of his doctoral work at Cambridge University, and IBM’s Deep Blue~\cite{campbell2002deep}, which famously defeated world champion Garry Kasparov in 1997. This landmark drew global attention and marked a turning point in public awareness of AI’s capabilities.

AI in games is commonly applied in creating game-playing agents, generating content, and modelling player behaviour~\cite{yannakakis2018artificial}. For game agents, algorithms such as Minimax~\cite{turing1953digital,russell2016artificial} have achieved superhuman performance across genres, including board games, strategy games, and first-person shooter games~\cite{yannakakis2018artificial}.
Since the early 1980s, games have used procedural content generation (PCG)~\cite{shaker2016procedural} techniques to dynamically create content. \textit{Rogue} (1980) and \textit{Elite} (1984) pioneered this approach by generating dungeons and galaxies from random seeds to ensure unique playthroughs. PCG has since been adopted in commercial games such as \textit{Diablo~III} (2012) and \textit{No Man’s Sky} (2016), enabling large, varied, and re-playable environments. AI techniques are also used to model player behaviour, enabling dynamic adaptation of gameplay by predicting preferences, emotions, or play styles from in-game interactions and physiological signals~\cite{yannakakis2013player}.

In the era of Generative AI, game development workflows have shifted significantly, enabling new creative possibilities.
Tools such as DALL-E and Stable Diffusion allow designers to rapidly generate high-quality visual assets, accelerating prototyping and reducing production overhead~\cite{vimpari2023adapt}. 
Park et al.~\cite{park2023generative} propose \textit{generative agents} that simulate believable human behaviour in interactive environments using LLMs. 
These agents autonomously interact, form relationships, and coordinate group activities, with applications in prototyping, narrative exploration, and dynamic NPC design.

\smallsection{Open-Source Game Development}
Scacchi~\cite{scacchi2004free} conducted an empirical study of free and open-source software (FOSS) development practices across multiple communities, highlighting how the game development community exemplifies key FOSS processes, such as informal requirements specification, decentralized version control, and layered meritocracy, which diverge significantly from traditional software engineering process models.
Pascarella et al.~\cite{pascarella2018video} conducted a mixed-methods study of 60 open-source projects to examine how game development differs from traditional software development. 
Their findings reveal notable differences in artifact types, developer roles, fault handling, and testing practices. 

\subsection{Github Issue Analysis}
GitHub is widely used in research for its popularity and accessible API~\cite{kalliamvakou2016depth}.
Introduced in 2009, GitHub Issues is an integrated system for reporting bugs, requesting features, asking questions, and managing tasks.

Bissyand{\'e} et al.~\cite{bissyande2013got} conducted the first large-scale study of GitHub issues, analyzing over 100{,}000 projects to examine issue reporting practices. 
They found that despite default availability, only a minority of projects actively use issue trackers. 
Issue activity correlates strongly with project size, team size, and developer popularity. Most issues are tagged as bug reports or feature requests, and distributed collaboration (e.g., via forks) increases user participation.

Lee et al.~\cite{tan2020first} conducted a large-scale study of over 70{,}000 GitHub issues labelled as ``good first issue” to examine their characteristics, onboarding value, and potential for automated detection. Their findings highlight how issue discussions reflect real-world developer practices and support the use of issue mining to study software collaboration dynamics.
Yang et al.~\cite{yang2023users} empirically studied 24{,}953 issues from 576 GitHub repositories linked to AI papers from top-tier conferences to understand common user challenges in open-source AI development. Using open card sorting, they developed a taxonomy of 13 issue types, with runtime errors and unclear instructions being the most frequent. They also found that while issue management features like labels and assignees are underused, they significantly improve closure rates.

While prior research has established foundational knowledge about AI techniques in games and practices in open-source development, few studies have investigated how recent GenAI technologies are reflected in developer workflows. Our study complements these efforts by focusing on issue discussions as a lens into practical GenAI-related challenges and practices, extending the literature with empirical insight into how developers engage with emerging AI capabilities in open-source game contexts.

\section{Research Questions}

This study is exploratory in nature; we aim to uncover patterns and themes rather than test predefined hypotheses. Our goal is to investigate the major  
challenges encountered by open-source developers working on games that incorporate AI technologies.  To achieve this, we analyze \emph{issue discussions} retrieved from project repositories and classify them into three categories: \emph{GenAI} (pertaining to the use of Generative AI), \emph{TradAI} (pertaining to the use of non-generative or traditional AI), and \emph{NonAI} (unrelated to AI).  Our immediate objective is to compare the types of development activities and challenges associated with each category, with particular attention to the novel concerns introduced by the integration of AI.  Ultimately, this work aims to advance understanding of the broader challenges associated with AI adoption in game development and to inform more effective and sustainable practices for leveraging both GenAI and TradAI technologies.

Recent advances in GenAI are beginning to transform the way games are designed, developed, and played. 
However, game development presents unique challenges for GenAI integration, due to its inherently interdisciplinary nature and high demands for narrative, aesthetic, and player experience coherence. 
Despite growing interest, there is limited empirical understanding of how developers are using GenAI in practice, particularly in open-source game development.
This study aims to bridge that gap by identifying what kinds of GenAI technologies are in use, what problems and opportunities they introduce, and how they differ from TradAI or NonAI development workflows. 
To guide our investigation into the impact of GenAI on open-source game development, we propose three research questions:

\begin{rqbox}{RQ1}
\rqonetext 
\end{rqbox}

\smallsection{Motivation} Recent advances in GenAI have introduced a wide range of tools and capabilities into game development, but it remains unclear how these technologies are being used in practice.

\smallsection{Goal} 
We aim to explore characteristics of how GenAI is utilized in open-source game development. 
This includes identifying the types of GenAI technologies mentioned, the tasks they support, and their modes of integration. 
By analyzing issue discussions, we seek to uncover patterns in how GenAI is referenced, applied, and embedded within development workflows.

\begin{rqbox}{RQ2}
\rqtwotext
\end{rqbox}

\smallsection{Motivation} While the adoption of GenAI in game development is growing, it is not yet clear what developers perceive as its most useful contributions, or where they encounter the most friction.

\smallsection{Goal} 
We aim to uncover the perceived 
challenges associated with integrating GenAI into open-source game development. 
By analyzing the content and type labels assigned during our issue classification process, as well as through thematic analysis~\cite{braun2006using} of GenAI-related issues, we seek to identify common pain points.

\begin{rqbox}{RQ3}
\rqthreetext
\end{rqbox}

\smallsection{Motivation} Although AI has long played a role in game development, the emergence of GenAI introduces new tools, practices, and challenges that may differ significantly from previous practices. To understand whether GenAI truly represents a shift in developer behaviour or simply extends existing paradigms, it is important to compare GenAI-related issues with those from other categories.

\smallsection{Goal} 
We aim to compare the characteristics and challenges of GenAI-related issues to those observed in TradAI and NonAI discussions. Here, we use \textit{characteristics} in the same sense as defined in RQ1, referring to the tools, tasks, and usage patterns developers describe in issue discussions.
By comparing issues across the three groups, we seek to identify whether GenAI introduces unique patterns in usage, developer concerns, or challenges not commonly found in other types of issues.

\section{Study Design}
To support our analysis, we collect a stratified sample of issues from open-source game repositories and categorize them into \emph{GenAI}, \emph{TradAI}, or \emph{NonAI}, as described below. 

\begin{figure*}[htbp]
\centerline{\includegraphics[width=\linewidth]{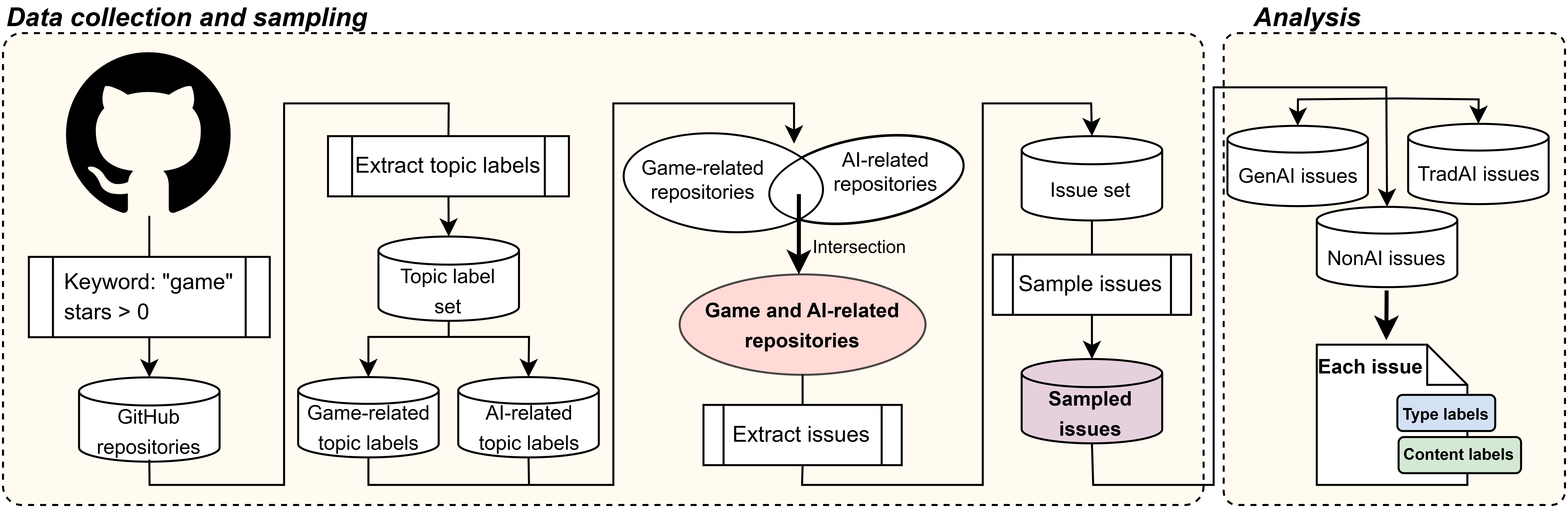}}
\caption{Overview of our study design}
\label{fig}
\end{figure*}

\subsection{Repository Selection and Filtering}

\smallsection{Target Repository Selection} The goal of this step is to identify open-source repositories that are relevant to both game development and AI, with particular emphasis on those involving GenAI technologies. 
We are planning to begin by querying GitHub for public repositories that contain the keyword \texttt{"game"} anywhere in their metadata (e.g., name, description, README) and have at least 1 star up to April 30th, 2025. This initial threshold is intentionally set to be minimal to maximize coverage and ensure that we collect a comprehensive set of repository topic labels. Starting with a stricter filter (e.g., requiring 10 or more stars) could risk excluding less popular but still thematically relevant repositories, potentially biasing the topic label pool. Due to GitHub’s 1,000-result limit per search, we aim to segment the query into 3--4 day intervals based on repository creation dates.

Next, we extract GitHub topic labels for each repository to construct an initial set of topics. 
To identify our final target repositories, those associated with both games and AI, we then carry out the following steps based on their assigned topic labels:
\begin{itemize}
    \item Collect all unique topic labels; exclude all repositories that do not contain any topic labels.
    
    \item To prioritize reviewable topics, we rank all topic labels by frequency and calculate how much additional repository coverage each new label provides. We stop adding topics once the marginal gain drops below 0.5\%, identifying an elbow point \textit{N} where further topics add little coverage. 
    
    \item Manually review the top \textit{N} topic labels to develop an initial understanding of which labels are associated with game development, AI, and GenAI.
    
    \item Starting from seed keywords (e.g., \texttt{game} for game-related topics, \texttt{artificial-intelligence} for AI-related topics, and \texttt{large-language-models} for GenAI), we will collect all repositories tagged with these topics and examine their co-occurring topic labels. 
	We iteratively expand our sets of game-related and AI-related topic labels by including newly discovered labels and repeating this process until no new topics are added.
    
    \item Classify a repository as game-related if it contains at least one topic label from the game-related set, and as AI-related if it contains at least one topic label from the AI-related set.
    
    \item Finally, we will take the intersection of the game-related and AI-related repositories. 
	This yields our final dataset of interest: open-source projects that lie at the intersection of games and AI.

\end{itemize}
The technique used to select the top N topics of the repositories provides a cutoff which balances review effort and coverage, allowing us to focus manual review on a manageable number of topics that collectively cover the majority of relevant repositories. For example, if the 201st topic adds $<$0.5\% repositories compared to the 200th, we set N = 200.

Note that we do not differentiate AI and GenAI into separate topic label sets at the data collection stage because many topic labels are ambiguous or overlapping, and GenAI is often represented as a subset of broader AI-related topics. Instead, this distinction is introduced later during issue-level classification, where we can make more accurate determinations based on contextual information.

We do not filter repositories by programming language, aiming to capture a diverse range of repositories. Since GenAI tools are often accessed via APIs, their use could be language-agnostic. However, we acknowledge that dominant languages may influence issue content, which we account for during qualitative analysis.

\smallsection{Repository Filtering} To ensure that we focus on active and relevant projects, we apply the following additional filters: 
\begin{itemize}
    \item Repositories must have at least 10 stars to reflect a minimum level of community interest. 
    \item Repositories must have at least 10 commits in 2024 or later
    to ensure active development and engagement.
    \item Repositories must not be forked projects to ensure there are no duplicates or near-duplicates in the data set.
    \item Repositories must not be educational projects, as these do not reflect real-world development practices we aim to study. 
	This filtering step is performed manually.
\end{itemize}
Following prior research~\cite{kalliamvakou2014promises,abdalkareem2017developers,jafari2021dependency}, we believe that these criteria and thresholds help prioritize repositories that are active, well-maintained, and broadly representative of real-world development practices.

\subsection{Issue Collection and Sampling}

We intend to collect all issues from our filtered target repositories, and for our analysis, we primarily focus on the issues that have non-empty labels, along with the following metadata: \emph{status} (i.e., open or closed), \emph{title}, \emph{body}, \emph{creation time}, \emph{last updated time}, \emph{closing time} (if applicable), \emph{number of comments}, and \emph{associated labels}.

Given the large number of issues and limited human resources, it is infeasible to manually inspect all of them. 
Therefore, we will extract a stratified random sample of these issues. 
The stratification is done per repository where issues are sampled proportionally based on the number of issues in each target repository. This ensures that large or highly active repositories do not dominate the sample.
Specifically, we will use a sample size formula for an unknown population~\cite{daniel2018biostatistics}, which ensures a 95\% confidence level with a margin of error of ±3\%. This yields a minimum sample size of approximately 1,000 issues, depending on the total size of the filtered dataset. 

Given the recent emergence of GenAI, if our initial manual classification, which will be described later, yields too few GenAI-related issues, we plan to leverage a large language model to assist in identifying additional candidates, building on prior research that has shown the potential of LLMs to assist manual annotation efforts in software engineering tasks~\cite{ahmed2025can}. Specifically, we plan to use DeepSeek, a performant open-access LLM with strong reasoning capabilities. We selected DeepSeek over alternatives like GPT-4 primarily due to the availability and cost-efficiency. To ensure classification quality, we will construct prompt templates that include: 1) a clear task instruction, 2) a definition of each category, and 3) the full issue content (title + body). We will apply few-shot prompting and enhance the classification process by reusing the initial manual labels (\emph{GenAI}, \emph{TradAI}, or \emph{NonAI}) and submitting the same classification task to Deepseek.
If the model achieves a sufficiently high agreement rate with our manual labels, we will then apply the tool to all issues with non-empty labels, ask it to identify those it considers GenAI-related, and all such flagged issues will still undergo manual verification before inclusion in the final dataset. 

We will also use the model’s full classification result to estimate the temporal distribution of issue categories, and to guide the scale of GenAI expansion so that it remains consistent with the estimated prevalence of GenAI, TradAI, and NonAI issues. 

\section{Execution Plan}

\subsection{Initial Issue Classification}

At the beginning of our study, we will perform a manual content-based classification by inspecting each issue and assigning it to one of three categories: \emph{GenAI}, \emph{TradAI}, or \emph{NonAI}. Each issue is reviewed by at least two annotators who independently examine the issue’s title and body.
Annotators are guided by a shared rubric that defines GenAI as referring to generative technologies such as large language models. TradAI refers to classical AI techniques such as decision trees and pathfinding algorithms. NonAI issues include all others, such as defects and new features.
After the independent annotation phase, we compute Cohen’s Kappa to assess inter-rater agreement. In cases of disagreement, the annotators meet to discuss and reach a consensus classification.

Note that classification is performed at the \textit{issue} level, not the repository level, since many repos may involve more than one category. If an issue discusses both, we assign the label based on the primary focus inferred from context. Ambiguous cases are resolved through team discussion and consensus, allowing us to capture GenAI adoption without overgeneralizing at the repository level.

\subsection{Open Card Sorting}
To address our research questions, we plan to perform open card sorting~\cite{breu2010information} on issues from the three groups. We will assign one or more labels to each issue based on the following two dimensions:
\begin{itemize}
    \item \textit{Type}: This refers to the intent or purpose of the issue, and answers the question: \textit{Why was this issue created?}
    \item \textit{Content}: This refers to the technical or functional focus of the issue, and answers the question: \textit{What area of the system does this issue relates to?}
\end{itemize}

For type labels, we intend to reference GitHub’s default issue label taxonomy as a starting point. This includes five predefined labels:

\begin{itemize}
    \item \textit{Bug report}: Indicates an unintended behaviour, malfunction, or defect in the system.
    \item \textit{Feature request}: Suggests adding new features or capabilities.
    \item \textit{Improvement}: Proposes an enhancement to an existing feature, logic, or user experience.
    \item \textit{Discussion}: Represents an open-ended conversation, brainstorming session, or request for clarification.
    \item \textit{Help wanted}: Signals a request for assistance or external contributions to resolve the issue.
\end{itemize}
We introduce new type labels during the card sorting process if none of the existing ones adequately capture the nature of an issue. We also allow multiple type labels to be assigned to a single issue if it reflects more than one intent or purpose.

For content labels, the complexity of issue content makes it difficult to rely on a fully predefined set. During the open card sorting process, we will assign multiple content labels to each issue to represent its high-level themes. These labels will include not only abstract concepts (e.g., prompt engineering), but also specific models or application names mentioned in the issue (e.g., ChatGPT) to more accurately capture their content.

To ensure reliability in the open card sorting process, similar to the initial issue classification, each issue is independently labelled by at least two annotators with respect to both type and content. During the pilot phase, we calculate inter-rater agreement (using Cohen’s Kappa) to assess consistency and refine label definitions. When disagreements arise, annotators review the issue collaboratively and resolve labels through discussion and consensus. The label set itself is refined iteratively, based on recurring patterns and ambiguity observed during annotation. 

\subsection{Thematic Analysis}
At the end of the open card sorting process, each issue will be assigned one or more types and content labels. These labels provide a high-level abstraction of the issue’s purpose and technical theme. 
While flat content labels provide a useful high-level overview, they often overlook the relationships between technical areas, models, and tasks. To better understand these relationships and build a more expressive hierarchical relationship for analysis, we conduct a thematic analysis to group and organize labels into a structured taxonomy.

We begin by reviewing the full set of content labels generated during the open card sorting phase. Through iterative refinement, we group semantically similar labels into broader categories. We also assign tool- and model-specific labels with the functional themes they support. The resulting structure can be interpreted as a hierarchical taxonomy that reflects both abstract capabilities and concrete implementations.

Our taxonomy provides a foundation for deeper content-level analysis in subsequent phases. 
It allows us to identify recurring patterns across projects, examine tool usage trends, and explore the structure of AI-related, NonAI-related, and especially GenAI-related workflows and integrations.

\section{Analysis Plan}

We now discuss the analysis that will be performed to address each of our
research questions.\\[-0.95em]

\noindent\textbf{RQ1:} \emph{\rqonetext}\\[-.95em]

\smallsection{Analysis}  
This question focuses on identifying what types of GenAI tools, supported tasks, and usage patterns are visible in developer discussions within open-source game repositories.
We will analyze the subset of issues labelled as GenAI-related. 
For each issue, we will examine the assigned content labels to identify the types of tools, tasks, and usage patterns discussed. 
We will aggregate label frequencies to determine which GenAI capabilities are most commonly referenced. 
We will also analyze co-occurrence patterns between content labels and specific models or tools (e.g., ChatGPT) to reveal integration trends. 
The results will be summarized using descriptive statistics, bar charts, and co-occurrence graphs derived from our taxonomy.\\[-.95em]

\noindent\textbf{RQ2:} \emph{\rqtwotext}\\[-.95em]

\smallsection{Analysis} 
This question aims to uncover the specific challenges and pain points developers encounter when adopting GenAI tools in open-source game development. 
Similar to RQ1, we will conduct a focused analysis of GenAI-related issues, using both the assigned type labels and open-coded themes. 
We will identify and group issues that reference the challenges. 
Thematic clustering will be performed based on recurring expressions and developer concerns in issue titles and bodies. 
Where appropriate, we will highlight illustrative issue examples to showcase specific pain points.\\[-.95em]

\noindent\textbf{RQ3:} \emph{\rqthreetext}\\[-.95em]

\smallsection{Analysis} 
We intend to compare GenAI-related issues with others to explore what makes their development distinct. 
We will first perform the same analysis on both TradAI-related and NonAI-related issues. 
We will then compare labelled issues across all groups by analyzing commonalities and differences in the distribution of type labels and content labels in RQ1, as well as challenge themes in RQ2. 
We will also assess whether GenAI-related issues tend to involve more tool-specific discussions or higher conceptual complexity.

\section{Threats To Validity}
We now outline potential threats to the validity of our study.\\[-.95em]

\smallsection{Internal Validity} Our repository filtering pipeline relies heavily on GitHub topic labels to identify game-related and AI-related projects. However, these labels are community-maintained and often inconsistently applied, which may result in the exclusion of relevant projects or inclusion of unrelated ones. We use stratified random sampling to select issues proportionally from each repository, but repositories with fewer issues or lower activity may be underrepresented, potentially biasing the dataset toward larger or more active projects. Additionally, when analyzing tool mentions, we assume that these references reflect actual tool usage. In practice, some mentions may be exploratory, comparative, or speculative rather than indicative of real integration. To address this, annotators are instructed to consider the context of each tool mention and, where possible, distinguish between exploratory discussions, comparative references, and concrete integration during manual labelling to reduce false positives.

\smallsection{Construct Validity} 
Our manual classification of issues into GenAI-related, TradAI-related, and NonAI groups relies on human judgment, introducing potential for misclassification, particularly when issue descriptions are vague or context-dependent. 
While we employ open card sorting with multiple annotators and track inter-rater agreement, labelling remains inherently subjective. 
The effectiveness of tools like DeepSeek in assisting classification also depends on the model’s ability to accurately interpret developer language, which may not always align with human understanding.
Finally, we assume that issue discussions reliably reflect developer concerns and experiences, but some challenges or benefits may go undocumented or be discussed elsewhere.

\smallsection{External Validity} Since our study focuses exclusively on public open-source repositories hosted on GitHub. This focus reflects a practical constraint: while our broader research goal is to understand how GenAI is used in game development, open-source projects provide accessible and analyzable data to support empirical investigation.
As such, the results may not extend to private repositories, commercial game development environments, or other platforms such as GitLab or Bitbucket. Moreover, our data collection is time-bounded, capturing repositories created up to April 30th, 2025. Given the rapid pace of advancement in GenAI tools, the technologies and practices observed may shift significantly over time. Finally, we implicitly focus on English-language repositories and issue discussions, which may underrepresent developer communities operating in other languages or regions, thereby limiting the global generalizability of our conclusions.

\section{Conclusion}

This study plan aims to explore how generative AI (\emph{GenAI}) is discussed and integrated in open-source game development by analyzing GitHub issue discussions. 
By classifying issues into \emph{GenAI}, traditional AI (\emph{TradAI}), and \emph{NonAI} categories, and applying structured labelling and thematic analysis, we seek to identify the tools, tasks, and challenges associated with GenAI usage.
By characterizing how GenAI is adopted and discussed in OSS game development, we hope to provide early empirical insights and inspire future research and tooling efforts at the intersection of AI and games.

\newpage

\bibliographystyle{plain}
\bibliography{ref}

\end{document}